\documentclass[12pt]{article}
\newcounter{sec}
\newlength{\dummysp}
\newcommand{\beqa}{\begin{eqnarray}}
\newcommand{\eeqa}{\end{eqnarray}}

\newcommand{\gappeq}{\mathrel{\rlap {\raise.5ex\hbox{$>$}}
{\lower.5ex\hbox{$\sim$}}}}
\newcommand{\lappeq}{\mathrel{\rlap{\raise.5ex\hbox{$<$}}
{\lower.5ex\hbox{$\sim$}}}}

\newcommand{\ben}{\begin{enumerate}}
\newcommand{\een}{\end{enumerate}}

\def\[{\left [}
\def\]{\right ]}
\def\({\left (}
\def\){\right )}
\newcommand{\beq}{\begin{eqnarray}}% can be used as {equation} or
\newcommand{\eeq}{\end{eqnarray}}

\begin{document}

\begin{titlepage}

\begin{center}

\hfill {\tt hep-th/0306204}\\

\vskip 1.2in

{\LARGE \bf Deconstructing KK Monopoles}

\vskip .4in

{\bf Erich Poppitz}

\vskip 0.2in

{\em Department of Physics, University of Toronto

Toronto, ON, M5S 1A7, Canada} 

\vskip 0.1in

 {\tt  poppitz@physics.utoronto.ca}

\end{center}

\vskip .5in
\begin{abstract}

\bigskip

We describe a procedure for finding Kaluza-Klein monopole solutions in deconstructed  four and five dimensional supersymmetric gauge theories. In the deconstruction of a  four dimensional theory, the KK monopoles are finite-action  solutions of the Euclidean equations of motion of the finite  lattice spacing theory.   The ``lattice" KK monopoles can be viewed as  constituents of continuum-limit  four dimensional instantons. In the five dimensional case, the  KK monopoles are static finite-energy stringlike configurations, wrapped and twisted around the compact direction, and can  similarly   be interpreted as 
constituents of five dimensional finite-energy gauge solitons. We discuss the quantum numbers and zero modes of the towers of deconstructed KK monopoles and their significance for  understanding   anomalies and  nonperturbative effects in deconstruction.
\end{abstract}
\end{titlepage}

\section{Introduction and summary}

Deconstruction was proposed  as  an ultraviolet regulator, or ultraviolet completion, of 5d and 6d gauge theories \cite{Arkani-Hamed:2001ca},\cite{Hill:2000mu}. 
The perturbative dynamics of the  4d ``quiver," or ``moose," product-group   theory (for early work on such theories, see  \cite{Georgi:au},\cite{Halpern:1975yj}) 
reproduces, in an appropriate  low-energy limit, the perturbative dynamics of the higher-dimensional theory. In the supersymmetric case, where exact nonperturbative results are available  \cite{Nekrasov:1996cz}, the nonperturbative effects in the compactified 5d theory are also captured by deconstruction \cite{Csaki:2001zx}. In the non-supersymmetric case, the use of deconstruction as a regulator is indispensable in the study of the small-instanton amplitude in  compactified 5d theories   \cite{Poppitz:2002ac}. 

The quiver theories of deconstruction appear naturally in string theory as the infrared-limit field theories on the worldvolume of D-branes at orbifold singularities \cite{Douglas:1996sw}. As was shown in \cite{Lykken:1997gy} (with a different motivation---the extra-dimensional nature of the Higgs-branch theory was not appreciated at the time),  an  orbifold of the Hanany-Witten brane configuration \cite{Hanany:1996ie}  gives rise to deconstruction of the simplest 5d supersymmetric theory  \cite{Arkani-Hamed:2001ca},\cite{Csaki:2001em}. 

The relation of deconstruction to branes on orbifolds has led to some interesting proposals. One is to use deconstruction to define the ill-understood 6d $(0,2)$, or $(0,1)$, theories   or  their compactifications  \cite{Rothstein:2001tu},\cite{Arkani-Hamed:2001ie}, see also \cite{Csaki:2001zx},\cite{Csaki:2002fy},\cite{Iqbal:2002ep},\cite{Constable:2002vt}. Another proposal
 is to use the deconstruction of 2d, 3d, or 4d  theories in terms of quiver quantum mechanics  or quiver matrix models  to give a lattice formulation of supersymmetric theories  \cite{Kaplan:2002wv},\cite{Cohen:2003xe}. The ``dynamical lattice"  preserves part of the continuum limit supersymmetry. In some cases, this helps  alleviate  the  fine-tuning required to  reach the supersymmetric critical point. For recent discussions  relevant for the practical feasibility of this proposal  see \cite{Giedt:2003ve},\cite{joel2}.

Whether we use deconstruction to give a lattice formulation  of supersymmetric theories in four or fewer dimensions, or as a definition of higher-dimensional theories, it is important to have an understanding of the symmetries  and  their anomalies. In the   5d and 6d cases, there are some interesting questions: for example, the 6d $(0,2)$ theories have an $SP(4)$ global symmetry with 't Hooft anomaly, predicted by anomaly inflow   to scale as $N^3$   with the rank of the theory  \cite{Harvey:1998bx},\cite{Freed:1998tg}, see also 
\cite{Intriligator:2000eq}. A microscopic understanding of this scaling is presently lacking;  one expects that a nonperturbative definition of the theory should be useful in  explaining it. The 5d theories, whose deconstruction was studied in refs.~\cite{Csaki:2001zx},\cite{Iqbal:2002ep}, are also known to have enhanced global symmetries,  predicted by string dualities \cite{Seiberg:1996bd},\cite{Intriligator:1997pq}, and a (workable) definition of the theory should exhibit them.

In the deconstruction of better-understood theories, where a weak-coupling continuum limit exists, the realization of the symmetries is significantly clearer. In accordance with general theorems \cite{Nielsen:1980rz},\cite{Nielsen:1981xu},  at finite lattice spacing the deconstructed theory violates the  global chiral symmetries. The symmetries are restored in the continuum limit. Their anomalies are well-understood in lattice perturbation theory \cite{Karsten:1980wd}. The relation of deconstruction to branes on orbifolds provides a geometric understanding of the chiral symmetry breaking at finite lattice spacing and of its  continuum-limit anomaly via string perturbation theory (or, with enough supersymmetry,  classical supergravity) \cite{Giedt:2003xr}.

In this paper, we continue the study of anomalies in the 3d $\rightarrow$ 4d deconstructed theories of  ref.~\cite{Giedt:2003xr}, where we discussed  the chiral $U(1)_R$ anomaly in both lattice and string perturbation theory (and supergravity). In the continuum theory, anomalies are also manifested as charge nonconservation in topologically nontrivial gauge field backgrounds. Our main goal here is to provide a similar understanding in the deconstructed theory and find the deconstructed version of a 4d instanton background.

We study the deconstruction of 4d ${\cal N}=2$ (eight supercharges) supersymmetric theory in terms of a 3d ${\cal N}=2$  (four supercharges) quiver gauge theory---the 4d Seiberg-Witten theory on a one-dimensional (deconstructed)  spatial lattice. 
We construct semiclassical field configurations of  finite Euclidean action, which, in the continuum limit, lead to nonconservation of the $U(1)_R$ charge. Our discussion is inspired by the continuum results for compactified 4d theories of refs.~\cite{Lee:1997vp},\cite{Lee:1998bb},\cite{Kraan:1998pm},\cite{Kraan:1998kp}; see also \cite{Davies:1999uw}. There,  instantons of the 4d theory are constructed from instantons of the 3d theory, i.e.,  Euclidean 't Hooft-Polyakov monopoles. In these references, the constituents of the 4d instanton are called   ``Kaluza-Klein monopoles."  The KK monopoles can be loosely thought of as excited (in the compact dimension) states of the purely-3d monopoles.

This paper is organized as follows. We begin, in section 2, by fixing our notation. We give the lagrangian and the bosonic equations of motion of the   3d ${\cal N}=2$ quiver theory. 

In sections 3.1, 3.2, we describe the construction of  KK monopoles as exact  solutions of the finite-$N$ deconstructed theory equations of motion. We should stress that the dynamical nature of the lattice (the fact that the lattice spacing is the expectation value of a field) is crucial to the ability to construct the KK monopoles as solutions of the finite-$N$ quiver theory; the procedure used to construct the KK monopole tower is rather general and can be applied in other deconstructed theories as well, for example, ones with a different amount of supersymmetry.
The KK monopoles break all of the explicit 3d ${\cal N} = 2$  supersymmetry, but at large $N$ are ``approximately BPS" (i.e., they become BPS in the infinite $N$ limit).  

In section 3.3, we discuss the  action,   quantum numbers, and  zero  modes of the deconstructed KK monopoles. 

In section 4,  we show that a field configuration of a monopole and an appropriate KK monopole has, in the large-$N$ limit, the action, quantum numbers, and zero modes of a 4d instanton. At infinite $N$, the configuration becomes supersymmetric  and the continuum analysis of  \cite{Lee:1998bb} applies, showing that an instanton configuration is exactly reproduced.  

Finally, we note that  while our main goal here is to study   anomalous charge nonconservation in our  example of 3d $\rightarrow$ 4d deconstruction,  our results are clearly relevant for the study of higher-dimensional supersymmetric theories and to the understanding of  their symmetries via deconstruction. For example, the particular quiver theory we study, when ``uplifted" from 3d to 4d, is the 4d $\rightarrow$ 5d deconstruction  \cite{Csaki:2001em},\cite{Csaki:2001zx}, of the simplest nonabelian supersymmetric  5d theory studied in \cite{Seiberg:1996bd}. The Euclidean monopole and KK monopole solutions uplift to static solutions of the 5d deconstructed theory. They represent magnetically charged strings wound and twisted around the extra dimension. Similar to the deconstruction of the 4d theory,  in the classical continuum limit they can be interpreted as constituents of  5d static finite energy solitons (which, in their turn, are simply the uplift of 4d Euclidean instanton solutions).  

The study of the nonperturbative states of the 5d theory is important, as these states couple to the 5d $U(1)$  topological current---the symmetry that (e.g., in the flavorless $SU(2)$ gauge theory) becomes enhanced at the fixed point  \cite{Seiberg:1996bd}. The quantization of the zero modes of these gauge solitons is a somewhat complicated problem, which  depends on the UV completion;   its study within deconstruction  is left for future work.
 
\section{Lagrangian and bosonic equations of motion}

In this section, we describe the theory we study. As mentioned in the introduction, this is  the  Seiberg-Witten theory [4d  ${\cal N} = 2$ super-Yang-Mills] on a one-dimensional spatial (deconstructed) lattice. In other words,  we consider  the 3d ${\cal N} = 2$ $SU(k)^N$
quiver gauge theory, with matter content that we describe using 
standard 4d ${\cal N} =1 $ superfield notation. The vector multiplets $V_i$, $i = 1, \ldots, N$, contain the 3d $SU(k)_i$ vector boson,  gaugino, and real adjoint scalar 
(the dimensional reduction of the third component of the 4d vector field). The ``link" superfields are chiral superfields, $Q_i$,  transforming as a bifundamental under the $i$-th and $i+1$-th gauge groups.

The tree-level Kaehler potential of the 3d ${\cal N} = 2$ $SU(k)^N$ quiver theory is:
\beq
\label{kahler}
K~ = ~{1 \over g_3^2} ~\sum\limits_{i=1}^N {\rm tr} ~Q_i^\dagger ~e^{V_i}~ Q_i~ e^{- V_{i+1}} 
\eeq
where $Q_i \rightarrow e^{\Lambda_i } Q_i e^{- \Lambda_{i+1}}$, $e^{V_i} \rightarrow e^{- \Lambda_i^\dagger}
 e^{V_i} e^{- \Lambda_i}$ under gauge transformations with   chiral gauge parameters $\Lambda_i$. To close the moose, we identify $N+ j \sim j$.
The $F$-term lagrangian is given by:
\beq
\label{super}
W ~=~ {1 \over 2  g_3^2}~ ~\sum\limits_{i=1}^N {\rm tr} ~W^\alpha (V_i) ~W_\alpha(V_i)~.
\eeq
We keep all fields to have canonical dimensions appropriate to 4d, so $g_3^2$ has dimension of mass. We are using generators in the fundamental, so that tr $T^a T^b = \delta^{ab}/2$ and $[T^a, T^b]= i f^{abc} T^c$. Further we will also need $g^{abc} =$tr($T^a T^b T^c) = {1\over 4} i f^{abc} + {1\over 2} d^{abc}$, with $d^{abc} = {\rm tr} T^a \{ T^b, T^c\}$.  For notational simplicity, in what follows, we will   only consider $SU(2)$, hence $d^{abc} = 0$, $f^{abc} = \epsilon^{abc}$. We let $m,n = 0,1,2$ denote 3d spacetime vector indices.

For latter convenience, we decompose the link field fluctuations into $v_j$ and $q_j^a$: 
\beq
\label{linkdefs}
Q_j ~=~ \sum\limits_{j = 1}^4 Q_j^A T^A~=~ v_j {\bf 1} +\sum\limits_{j = 1}^3 q_j^a ~T^a ~~ (T^4 \equiv {\bf 1})~,
\eeq
in terms of which 
the covariant derivatives are:
\beqa
\label{covtderivative}
D^m Q_j &=& \partial^m Q_i + {i \over 2} ~A_j^m Q_j - {i \over 2} ~Q_j A^m_{j + 1}~,\nonumber \\
&=& \partial^m v_j + {i\over 2} (A_{m j}  - A_{m j+1}) v_j  + \partial^m q_j + {i \over 2} (A_{m j} q_j - q_j A_{m j + 1})~. 
\eeqa
The covariant derivatives of the bosons $A^m_j$ and $\phi_j$ in the adjoint vector supermultiplets are: 
\beqa
\label{cov2}
D^m \phi_j &=& \partial^m \phi_j + {i\over 2}~ [A^m_{ j}, \phi_j]~, \nonumber \\
F_j^{mn} &=&  \partial^m A_j^n - \partial^n A_j^m + {i \over 2} ~[A_j^m, A_j^n ] ~.
\eeqa
In terms of (\ref{covtderivative}), (\ref{cov2}), the bosonic lagrangian (for completeness, we note that the fermionic part of the lagrangian was given, using the same notation, in ref.~\cite{Giedt:2003xr})  is: 
\beqa
\label{lagrbos}
g_3^2 ~ L_{bos} ~=~ &-&  \sum\limits_{j = 1}^N \left\{ ~{1\over 2} {\rm tr}~ F_j^{mn} F_{mn \; j} + {\rm tr}~D_m \phi_j D^m \phi_j  + {\rm tr} (D_m Q_j)^\dagger D^m Q_j ~\right\} \nonumber \\
&-& \sum\limits_{j = 1}^N ~{1\over 4}~{\rm tr} |\phi_j Q_j - Q_j \phi_{j+1}|^2 - \sum\limits_{j = 1}^N ~{1 \over 8} \left(
{\rm tr}( Q_j^\dagger T^a Q_j - Q_{j-1} T^a Q_{j-1}^\dagger) \right)^2 ~. ~~~~~~~~~~~  
\eeq 
The  bosonic equations of motion are as follows.
The 3d gauge field $A^m_j$ at the $j$-th site obeys: 
\beq
\label{eqngauge}
(D_m F^{mk}_j)^a - {1 \over 2} ~\epsilon^{abc} (D^k \phi_j)^b \phi_j^c +\left[ {i \over 2}~ {\rm tr}\left( Q_j^\dagger~ T^a ~D^k Q_j - 
T^a ~Q_{j -1}^\dagger ~D^k Q_{j-1} \right) +{\rm h.c.} \right] = 0~.
\eeq

{\flushleft T}he equation of motion of the scalar $\phi_j^a$ in the 3d ${\cal {N}}=2$ vector multiplet  is:
\beqa
\label{eqnscalar} 
(D_m D^m \phi_j)^a  \hspace{12cm} \\
- {1  \over 4} {\rm tr} \left[  \left( T^a Q_j ( Q_j^\dagger \phi_j - \phi_{j+1} Q_j^\dagger) - Q_{j-1} ~T^a(Q_{j-1}^\dagger \phi_{ j-1} - \phi_j Q_{j-1}^\dagger) \right) + {\rm h.c.} \right] =0 ~. \nonumber
\eeqa

{\flushleft F}inally,  the link fields  $Q_j$, with $T^A$ defined in eqn.~(\ref{linkdefs}) (a sum over $a = 1,2,3$ is understood) obey:
\beqa
\label{eqnlink}
{\rm tr}\left[ T^A \partial_m (D^m Q_j)^\dagger - {i \over 2} \left( A_{m j} T^A - T^A A_{m j+1}\right) (D^m Q_j)^\dagger 
-{1\over 4} \left( \phi_j T^A - T^A \phi_{ j+1}\right) \left(Q_j^\dagger \phi_j - \phi_{j+1} Q_j^\dagger\right)  ~~~~~\right. \nonumber \\
 \left. - {1\over 4}  Q_j^\dagger ~T^a T^A ~{\rm tr} \left( Q_j^\dagger T^a Q_j - Q_{j-1} ~T^a Q_{j-1}^\dagger \right)+ {1\over 4}  Q_j ~T^A T^a~  {\rm tr} \left( Q_{j+1}^\dagger ~T^a Q_{j+1} - Q_j T^a Q_j^\dagger \right)\right] = 0~ ~~~~~~~\eeqa

One consequence of the above equations of motion follows from considering ``diagonal,"  $j$-independent, field configurations,  such that $Q \equiv Q_1 = Q_2 = ...=Q_N$, $A^m \equiv A^m_1 =  ... = A^m_N$, and $\phi \equiv \phi_1 =...=\phi_N$. 

For diagonal field configurations,  the equations of motion (\ref{eqngauge}-\ref{eqnlink}) of the quiver theory reduce to the equations of motion of a 3d ${\cal N}=4$ $SU(2)$ theory, with vector field $A^m$ and three real scalar adjoints (Re$q$, Im$q$,  $\phi$) forming a triplet under the $SU(2)$ global symmetry of the 3d ${\cal N} = 4$ theory (here $q$ is defined as in eqn.~(\ref{linkdefs})),
along with the decoupled free 3d Laplace equation for $v$ ($= v_1 = v_2 = ... = v_N$). Thus, any classical solution of the 3d ${\cal N}=4$ theory can be  promoted to a solution of the deconstructed theory by diagonally embedding it in the $SU(2)^N$ ${\cal N} = 2$ quiver theory.

Finally, we note that the equations for the 4d $\rightarrow$ 5d deconstruction can be obtained from eqns.~(\ref{eqngauge}-\ref{eqnlink}) without any work: one simply replaces $m,n \rightarrow \mu, \nu$, allowing now $\mu, \nu = 0, 1,2,3$ and erases all terms involving the fields $\phi_i$.  Much of what we say will hold for 4d $\rightarrow$ 5d deconstruction. The Euclidean finite action solutions we study in the 3d$ \rightarrow$ 4d deconstruction are replaced by time-independent solutions of finite energy in the 4d $\rightarrow$ 5d  case, as mentioned in the introduction.

\section{Instantons on the diagonal Coulomb branch}

We will study physics along the Higgs branch of the $SU(2)^N$ theory, where the quiver theory has  a regular lattice interpretation.  The link fields have diagonal expectation values: 
\beq
\label{higgsbranch}
\langle Q_j \rangle = v,  ~{\rm independent ~of} ~j~,
\eeq
which break $SU(2)^N$ down to the diagonal $SU(2)_D$.
In the decomposition (\ref{linkdefs}), the fields $v_i$ represent local fluctuations of the lattice spacing $v$. We assume that these are frozen due to, e.g., $D$-terms of the gauged $U(1)$ factors (see section 3.1).  

Along the diagonal Higgs branch (\ref{higgsbranch}), the classical continuum limit is reached by taking $N \rightarrow \infty$, $v \rightarrow \infty$, with the size of the compact dimension  $\pi R = N/v$ kept fixed; the large volume limit can then also be taken. The  quoted relation between the radius and the parameters of the deconstructed theory, as well as the  quantum continuum limit are discussed (in our present notation) in detail in \cite{Giedt:2003xr}.
In the continuum limit,  the equations of motion (\ref{eqngauge}-\ref{eqnlink}) reproduce the continuum equations of motion of the 4d ${\cal N} = 2$ theory. Thus, a  classical 
solution of the continuum theory should have a counterpart in the deconstructed theory: there should exist at least an approximate, at large-$N$, solution of the deconstructed theory equations of motion that becomes an exact solution in the continuum limit. 

In a continuum 4d compactified theory the finite Euclidean action  BPS solutions are the 4d  (periodic) instanton, as well as monopoles  on the 4d Coulomb branch,  or  on the Coulomb branch with a gauge field Wilson line turned on. It has been argued that the 4d instanton is composed of a  monopole and a certain KK monopole \cite{Lee:1997vp}. We  find a similar description in deconstruction. We will construct the constituents of a 4d instanton as explicit solutions of the quiver theory equations of motion at finite $N$.

It is natural to begin building finite action Euclidean solutions of the  continuum limit theory from finite action solutions of the deconstructed theory. In the following sections, we will explore the finite-$N$ instanton solutions of the $SU(2)^N$ theory and discuss their continuum counterparts.

\subsection{The diagonal monopole}

We begin by studying site-independent (diagonal) solutions.
As already mentioned, along the Higgs branch (\ref{higgsbranch}), the  theory of the zero modes
 is a 3d ${\cal {N}}=4$ theory (considering a restriction to $j$-independent fluctuations is  
consistent on the Higgs branch, where there is a finite cost of energy to $j$-dependent fluctuations).
For this enhanced supersymmetry of  the zero mode sector to be exact, the gauge group has to be $U(2)^N$ rather than $SU(2)^N$. We have avoided indicating this explicitly solely in order to not clutter notation. Vanishing of the $D$-terms of the $U(1)^N$ gauge groups then require that the vevs of neighboring $Q_j$ be equal, i.e. the lattice spacing be uniform. The diagonal $U(1)$ ${\cal {N}}=2$ vector multiplet combines then with the diagonal $j$-independent vev (\ref{higgsbranch}) to form a  3d ${\cal {N}}=4$  vector multiplet.

The Euclidean zero mode theory has an $SU(2)_E \times SU(2)_N \times SU(2)_R$ global symmetry, where $SU(2)_E$ is the Euclidean space-time symmetry. 
The bosons ($\phi^a, {\rm Re} q^a, {\rm Im} q^a$) form a triplet under $SU(2)_N$. 
The zero-mode fermions of the vector and chiral multiplets form a doublet under $SU(2)_R$. The eight supercharges form a   $({\bf 2,2,2})$ multiplet under the global symmetry. The zero mode theory is the $R \rightarrow 0$ limit of a 4d ${\cal {N}}=2$ theory (whose compactification was studied in \cite{Seiberg:1996nz}) with fixed coupling $g^2_{3, D} = g_4^2/(2 \pi R)=  g^2_3/N$. It thus inherits all  classical solutions of the latter that have finite action in the above zero-radius limit. 

The zero mode $SU(2)_D$ theory has a   Coulomb branch of real dimension three, parameterized by the expectation values of the $SU(2)_N$ triplet of adjoint scalars ($\phi^a, {\rm Re} q^a, {\rm Im} q^a$). 
In what follows, we will consider a point along the Coulomb branch where  $q^a$ has an expectation value:
\beq
\label{branch}
\langle Q_i \rangle = v + i~ 2  \eta~ T^3~,
\eeq
where, generally,  $\eta = \eta^\prime + i \eta^{\prime\prime}$; we can always choose real inverse lattice spacing $v$. In the zero mode sector, one can always use $SU(2)_N$ to rotate the vev of $q$ into any desired direction, e.g. $\eta^\prime$. (In the full theory, however, these directions are not equivalent: the $\eta^\prime$ expectation value corresponds to turning on a Wilson line for the $4d$ gauge field, while $\eta^{\prime\prime}$ is a vev on the 4d Coulomb branch; this can be seen   from the large-$N$ identifications of the fields with those of a 4d ${\cal N} = 2$ theory, see \cite{Giedt:2003xr}.)

For further use, it is instructive to write explicitly the instanton solution of the zero mode $SU(2)_D$ theory in terms of the fields of the full theory.  The $SU(2)_D$ instanton is simply the Bogomol'nyi--Prasad--Sommerfield monopole solution
with $q^a$ playing the role of the scalar triplet field: 
\beqa
\label{BPSdiagonalsolution}
A^{m a}_j &=& {1 + \chi(r) \over r^2}~ \epsilon^{m a k} x^k~, \\
Q_j &=& v\: {\bf 1} + i ~e^{i \: {\rm arg} ( \eta) }~ {f(r) \over r^2} ~T^a x^a~, ~~ ~ (j = 1, \ldots, N)~,\nonumber
\eeqa
where $r$ is the Euclidean radius vector and:
\beqa
\label{solution}
f(r) &=& - 2 +  2~ r |\eta|~ {\rm coth} \; r |\eta|~, \nonumber \\
\chi(r) &=&1 -  2~ r  |\eta| ~ {\rm  csch} \: r |\eta|~.
\eeqa
For $\eta^{\prime \prime} = 0$ the functions $f(r)$ and $\chi(r)$ can be obtained by solving the self-dual BPS condition  $B_m^a = (D_m {\rm Im} q)^a$, with $q^a$ of eqn.~(\ref{linkdefs}). For nonvanishing  $\eta^{\prime\prime}$ the use of  $SU(2)_N$ gives rise to the  arg($\eta$) phase factor in (\ref{BPSdiagonalsolution}). For explicit expressions for $B_m^a$ and $(D_m {\rm Im} q)^a$, see footnote after eqn.~(\ref{Q}).
The perhaps unconventional factors of 2 in (\ref{solution}) (and further in the paper, e.g., an extra factor of 4 in the monopole and instanton actions)  are because of our convention for the gauge transformation law $A \rightarrow U A U^\dagger - 2 i U d U^\dagger$, see section 2.\footnote{A conventional normalization of the action (\ref{lagrbos}), as well as  ``normal" gauge transformations, can be achieved in terms of new fields and coupling, given by the redefinition:
$g_3 = 2 \tilde{g}_3$, $A_m = 2 \tilde{A}_m$, and $q = 2 \tilde{q}$. In particular, this leads to the conventional solution, instead of (\ref{solution}) [without the various factors of 2], and changes
 the factor of $16 \pi |\eta|$ in eqn.~(\ref{diagonalaction}) to the usual  $4 \pi |\eta|$. Although this will plague us till the very end, we will stick with our present normalization and will recall it when necessary.}
The diagonal solution in (\ref{BPSdiagonalsolution}) is written in hedgehog gauge. The  expectation value of $|q|$ at infinity  is  $2 |\eta|$ (and points in the third isospin direction, if rotated into the string gauge).  

The Euclidean action of the diagonal instanton  (\ref{BPSdiagonalsolution}) is easily seen to be (to compute the action, we are free to rotate the solution into the Im$q$ direction):
\beq
\label{diagonalaction}
S_{diag.} =  {N \over g_3^2} \int d^3 x ~ B_m^a (D_m {\rm Im} q)^a  = {N \over g_3^2} \: 16  \pi   |\eta| = {2 \pi R \over g_4^2}~ 16 \pi  |\eta|~.
\eeq
In (\ref{diagonalaction}), we used the BPS condition, Gauss' law,   the asymptotics of the fields at infinity, the ``bare" coupling matching relation,  $g_4^2 = 2 g_3^2/v$,    and  the already quoted $N v^{-1} = \pi R$, see \cite{Giedt:2003xr}. 

The continuum theory 
interpretation of the diagonal solution with $|{\rm Im} q\vert_{\infty}| = 2 |\eta|$ is as follows: it is the
Euclidean BPS monopole,  independent on the $S^1$ coordinate,  of the 4d ${\cal N} =2$ theory on $R^3 \times S^1$ on the 
Coulomb branch with nonvanishing Wilson line  of the gauge field. 

To define the magnetic charge of the solution with respect to the single unbroken $U(1)$, we follow the continuum construction of \cite{Gross:1980br}. 
We begin by defining a unitary ``Wilson line"  $\Omega_j$. 
Over almost all of field space we can  use  the polar decomposition of the bifundamental,  $Q_j = P_j ~{\cal U}_j$, with $P_j$---the positive matrix $P_j = \sqrt{ Q_j Q_j^\dagger}$,  to first define a unitary link field 
${\cal U}_j = P_j^{-1} ~Q_j$. 
The Wilson line $\Omega_j$ is then defined as: 
\beq
\label{wilsonline}
\Omega_j  \; = \; {\cal U}_j \; {\cal U}_{j+1} \ldots \;  {\cal U}_{j-1}~.
\eeq
The magnetic charge of the solution is: 
\beq
\label{magneticcharge}
q_{\alpha; j}  ={1 \over 4 \pi}~  \int\limits_{S^2_\infty} {d^2 \vec{\sigma}} ~{\rm tr} \left({\cal P}_{\Omega_{j}, \alpha} \vec{B_j} \right)~,
\eeq
where $B_{m ~j} = {1\over 2} ~\epsilon_{m k l} F^{k l}_j$ is the magnetic field  at the $j$-th site and ${\cal P}_{\Omega_{j; \alpha}}$ is the projector on the $\alpha$-th eigenvalue of  $\Omega_j$:
\beq
\label{projector}
{\cal P}_{\Omega_{j; \alpha}} = {1\over 2 \pi i} ~ \int\limits_{C_\alpha}   d z~{1  \over z - \Omega_{j}}~,
\eeq
with  integrals taken over small contours $C_{\alpha}$ encircling the corresponding eigenvalue.
The eigenvalues of $\Omega_j$ do not depend on $j$ (the characteristic equation only 
contains traces of powers of the Wilson line, which do not depend on $j$). The magnetic charge (\ref{magneticcharge}) measures the winding of   $\Omega_j$ at infinity and is thus also $j$-independent [see appendix B of \cite{Gross:1980br}].
Note that the magnetic charge (\ref{magneticcharge}) is  $U(2)^N$  invariant and that  $\sum_\alpha {\cal P}_{\Omega_j, \alpha} = {\bf 1}$ together with tracelessness of $\vec{B}_j$ implies  $\sum_\alpha q_{\alpha; j} = 0$. Thus, for an $SU(k)$ gauge group we would have $k -1$ independent charges, while for $SU(2)$ we need to make a choice of eigenvalue. 

At Euclidean infinity, the diagonal solution (\ref{BPSdiagonalsolution}) has $W_j = v^N (1 + i \sigma^a \hat{x}^a ~{ |\eta| \over   v})^N$
with  $\hat{x^a}$  denoting a unit vector. At large $N$ and fixed $\pi R = N/v$, this gives
$\Omega_j \simeq \exp( i \sigma^a \hat{x}^a ~|\eta|~   \pi R)$ with eigenvalues $e^{\pm i |\eta|    \pi R}$  and corresponding projectors (\ref{projector}). Finally, at infinity $r^2 B_j^{ma} \rightarrow 2 \hat{x}^m \hat{x}^a$ (this  follows from the explicit formulae from the footnote after eqn.~(\ref{Q})).  Hence,  from   (\ref{magneticcharge}, \ref{projector}) we obtain
$q_\pm = \pm 1$ for the diagonal monopole. 
The $U(2)^N$ invariant expression for the magnetic charge, eqn.~(\ref{magneticcharge}), will be useful in the following sections.

\subsection{The  KK monopole}

In the continuum theory, there exists another class of solutions on the Coulomb branch with nonvanishing Wilson line: the twisted instanton solutions, which depend  on the $S^1$ coordinate. They are obtained from the $S^1$--independent solution by applying  appropriate large gauge transformations. We continue by  describing this ``KK monopole" solution in   deconstruction.

The diagonal instanton   (\ref{BPSdiagonalsolution}) is a solution of the quiver theory equations of motion (\ref{eqngauge}-\ref{eqnlink}), which are invariant under  the $U(2)^N$ global counterparts of the $U(2)^N$ gauge transformations. Thus, a  global  $U(2)^N$  transformation maps solutions into solutions. However, it  also changes  the boundary conditions on the   Higgs/Coulomb branch. Thus, a $U(2)^N$-transformed solution is, generally,  a solution in a different vacuum, where the theory may not even have a regular lattice interpretation.
The global  $U(2)^N$ symmetry can nevertheless be used to generate new solutions in the same vacuum,  as we  now describe.
\begin{enumerate}
\item
The first step is to  identify the  $U(2)^N$ global transformations that preserve the structure of the Coulomb branch vacuum (\ref{branch}):
\beq
\label{vacuum}
\langle Q_j\rangle = v ~{\bf 1}  + i ~ 2 \eta~ T^3~,~~{\rm for~ all}~ j = 1, \ldots, N~.
\eeq
Under a general $U(2)^N$ transformation, which acts on the link fields as $Q_j \rightarrow U_j Q_j U_{j+1}^\dagger$, the vevs 
(\ref{vacuum}) become:
\beq
\label{vacuum1}
U_j~ \langle Q_j \rangle~ U_{j+1}^\dagger~=~ v ~U_j U_{j+1}^\dagger + i ~2  \eta ~ U_j T^3 U_{j+1}^\dagger~. 
\eeq
We want   to generate solutions  of the deconstructed theory along the Higgs branch, where the theory has a regular lattice interpretation,  with $j$-independent  inverse lattice spacing.  Thus, 
we demand that the product $U_j U_{j+1}^\dagger$,   hence the inverse lattice spacings (the first term on the r.h.s. of (\ref{vacuum1})) be $j$-independent. We  also demand that the point on the diagonal Coulomb branch (the  term $ \sim \eta T^3$ in (\ref{vacuum1})) is mapped to a point on the Coulomb branch, but not necessarily with the same vev. In other words,  the $j$-dependent $U(2)$ matrices $U_j$ (i.e., the ones  that generate the solutions of interest to us) should be such that the vacuum $\langle Q_j \rangle $ of eqn.~(\ref{vacuum}) gets mapped to a new site-independent vacuum with expectation values $\langle Q^U_j \rangle$: 
\beq
\label{vacuum3}
\langle Q^U_j \rangle = v^U + i ~ 2 \eta^U ~T^3~  \equiv  ~ U_j~ \left( ~v ~+~ i ~2 \eta ~T^3~\right)~ U_{j+1}^\dagger ~  , ~~{\rm for~ all}~ j = 1, \ldots, N~, 
\eeq
where $v^U$ and $\eta^U$ are  $j$-independent. Periodicity of the moose theory also requires $U_{j+N} = U_j$. The   $j$-dependent global $U_j$ transformations  obeying (\ref{vacuum3})  will be constructed below.
\item 
Next, having enumerated  the possible transformations $\left\{U_j ;  j = 1, \ldots,N\right\}$ that obey (\ref{vacuum3}), and having  found the corresponding $v^U$ and $\eta^U$, we can construct  the $j$-independent diagonal monopole solution of the previous section, in the vacuum   given by $v^U$ and $\eta^U$ of the  r.h.s. of eqn.~(\ref{vacuum3}). We can now 
convert this diagonal solution in the $v^U, \eta^U$ vacuum to a solution in the $v, \eta$ vacuum (\ref{vacuum}) by
performing a $U(2)^N$ transformation with the $\left\{U_j^{-1} ;  j = 1, \ldots,N\right\}$.  This new solution is   
a $j$-dependent solution in the vacuum given by $v$ and $\eta$ (\ref{vacuum}) and is, as we will show later in this section, the deconstructed counterpart of the ``KK monopole" solution of the compactified 4d gauge theory  \cite{Lee:1997vp}--\cite{Kraan:1998kp}.
\end{enumerate}

As outlined\footnote{The reader interested only in the final form of the solutions should jump to the end of this section, eqns.~(\ref{stringdiagonal}--\ref{KKgauge}), and to section 3.3 for a discussion of the KK monopole tower's quantum numbers.}  in step 1. above, we now begin the construction of the KK monopole solutions by finding the $U_j$ that obey (\ref{vacuum3}).  We rewrite (\ref{vacuum3}) as follows:  
\beq
\label{vacuum4}
U_{j+1} =   B^{-1} ~U_j~A , ~~ {\rm hence} ~~ U_{k+1} = (B^{-1})^k ~U_1~A^k ,
\eeq
where $A$ and $B$ are defined as:
\beq
\label{AB}
A  &\equiv& v   + i~2  \eta ~T^3 , ~\\ 
B  &\equiv& v^U  + i ~2 \eta^U~T^3 .~\nonumber
\eeq
If $U_1$ was arbitrary, periodicity   $U_1 = U_{N+1}$ would require that
 $(B^{-1})^N = e^{i \alpha}~{\bf 1}$ and $A^{N} = e^{ - i \alpha} ~{\bf 1}$. Evidently,   for an arbitrary $U_1$,
at an  arbitrary point on the Coulomb branch, $v, \eta$, these conditions can not be satisfied.  

Then, it is easy to see that, in order for (\ref{vacuum4}) with $A$ and $B$ of eqn.~(\ref{AB}) and with $U_{N+1} = U_1$ to hold for an arbitrary Coulomb branch vev, 
 $U_1$ has to take either of the two forms:  \beqa
\label{U1}
U_1^{I} = \left( \begin{array}{cc} e^{i \alpha}& 0 \\ 0 & e^{- i \alpha} \end{array} \right) ~ ~{\rm or}  ~~~
U_1^{II} = \left( \begin{array}{cc} 0 & e^{i  \alpha} \\ - e^{- i \alpha} & 0 \end{array} \right)~,
\eeqa
where $\alpha$ is an arbitrary phase; we also neglected an arbitrary undetermined overall $U(1)$ phase in $U_1$. 
Note that the two allowed forms (\ref{U1}) of $U_1^{I,II}$  are related by a discrete $ SU(2)$ transformation:  $U_1^{II} = U_1^{I} ~i \sigma^2$.
The effect of $i \sigma^2$ on the vacuum (\ref{vacuum}) is to flip  the sign of $\eta$.

We note also that $U_1^{I} = e^{i \alpha \sigma^3}$ is a site-independent global gauge transformation which does not affect the classical solutions: $\alpha$ is, in fact, one of the four  bosonic zero modes of the KK monopole, along with its position in 3d Euclidean space.  In what follows, we will only  consider global transformations (\ref{vacuum4}) with $U_1  = {\bf 1}$. The modifications  for $U_1 = i \sigma^2$   will be remembered when needed.
 
To construct the form of the $j$-dependent global transformations $U_j$ that preserve the regular lattice structure, we first note that $U_1^I$ commutes   with $A$ and $B$. Then,  the periodicity condition from (\ref{vacuum4}) is simply that $(B^{-1} A)^N = {\bf 1} $. 
Hence $B^{-1}  A = e^{i  \pi l/N} \Xi^{(l)}$, where $\Xi^{(l)}$ is an $SU(2)$ matrix
obeying $(\Xi^{(l)})^N = (-1)^l~ {\bf 1}$ and $l$ is an  integer, defined  $mod \; 2N$; see, however the comment after eqn.~(\ref{Utheory}). 
Any such matrix can be written as $\Xi^{(l)} = e^{ i \vec{m} \cdot \vec{T} 2 \pi l/N}$  for some unit vector $\vec{m}$; recall that $\vec{T} = \vec{\tau}/2$. 
 Thus, we find that:
\beqa
\label{B}
U_{k+1}^l =  e^{i {\pi l \over N} k } ~e^{i {2 \pi l \over N}  k \vec{m} \cdot \vec{T}} ~U_1~,  ~~ B~= ~e^{- i {\pi l\over N}  } ~A~ e^{- i  {2 \pi l \over N}   \vec{m} \cdot \vec{T}}  ~.  
\eeq
 Note that the $SU(2)$ part of the transformation (\ref{B})
is antiperiodic (in ``theory space") for odd $l$.
The second equation in (\ref{B}) along with eqn.~(\ref{AB}) implies that $\vec{m}$ must be along $T^3$, i.e. $\vec{m} = (0,0,1)$. 
Now we can find the expressions for  $v^U$ and $\eta^U$, which we label by $l$ and denote, from now on,  by $v^l$, $\eta^l$: 
\beqa
\label{vU}
v^{l}&=&e^{- i {\pi l\over N}}  \left( v \cos {\pi l \over N} + \eta  \sin {\pi l \over N} \right) \simeq ~v~, \\
\eta^{l}&=& ~e^{- i {\pi l \over N}} \left( \eta \cos{\pi l \over N} -  v \sin{\pi l \over N} \right) \simeq  \eta - { \pi v  \over N} ~l   =  \eta - {l\over R} ~. \nonumber 
\eeqa
We have also indicated the expressions for $v^l$ and $\eta^l$   near $l = 0$: 
$l \ll N$, $\eta \ll v$, $ \pi R = N/v$---fixed (we could also consider $l \sim N$; see below). The transformation law for $\eta$ shows that the global $U(2)^N$ transformations (\ref{B}) that preserve the $SU(2)_D$ Coulomb branch structure are the deconstructed version of
the large gauge transformations shifting the value of the Wilson line $\eta^\prime$ (note that in eqn.~(\ref{vU}) both $\eta, v$ and $\eta^l, v^l$ can be complex; however, if $\eta$ and $v$ are real, so are the large-$N$ limits of $\eta^l$ and $v^l$ as the r.h.s. shows).

To summarize, the ``theory space" dependent global transformation has the  form:
\beq
\label{Utheory}
U_{k+1}^l = e^{i {\pi l \over N} k } ~e^{i {  \pi l \over N}  k \sigma^3} ~U_1; ~ ~l, k  = 0 \ldots  N~.
\eeq
Recall also that we are  allowed to perform transformations with $U_1 = i \sigma^2$. For such $U_1$, as is easy to see tracing our previous derivation,  the relation between $v^l, \eta^l$ and $v, \eta$  gets modified by $\eta$ having the opposite sign on the r.h.s. of eqn.~(\ref{vU}).

Another important point indicated in eqn.~(\ref{Utheory}) is that the physically indistinguishable values of $l$ are defined $ mod \; N$. This is clear from the fact that $U_{k+1}^{l + N} = U_{k + 1}^l$, 
$v^{l + N} = v^l$, and $\eta^{l + N} = \eta^l$, as implied by (\ref{vU}, \ref{Utheory}).

The vacua described by $v, \eta$ and $v^l, \eta^l$ related by (\ref{vU}) [or for $U_1 = i \sigma^2$, by (\ref{vU}) with $\eta \rightarrow - \eta$] are physically equivalent since they are related by   discrete global symmetries. This is best visualized in the brane picture: the discrete global transformations---which are the deconstructed analogue of the large gauge transformations of the continuum limit theory---simply correspond to redrawing the wedges of the orbifold, or, equivalently, to different pairing of D-branes. We  will not  need this interpretation here and refer the reader for  detailed explanation to section III.C of ref.~\cite{Csaki:2001zx}.

Now that we have enumerated all global $U(2)^N$ transformations that preserve the regular lattice structure and the diagonal Coulomb branch, 
we are ready to construct the KK monopole solutions we are interested in. As described in the beginning of this section, in step 2., we start with the diagonal solution (\ref{BPSdiagonalsolution}) in the vacuum $v^l , \eta^l$ of eqn.~(\ref{vU}). Next,  we perform a $U(2)^N$ transformation on the solution by 
$\left\{ (U^l_j)^{-1}, j = 1, \ldots N \right\}$, with the $U^l_j$ given in eqn.~({\ref{Utheory}). This transformation maps the solution in the vacuum $v^l, \eta^l$  into a solution in the vacuum $v, \eta$. Since the action is gauge invariant, it is still given by 
(\ref{SKK}) (but with $\eta$ replaced by $\eta^l$). The new solution in the $v, \eta$ vacuum is, however, $j$-dependent, since the transformations $U_j$ do not commute with (\ref{BPSdiagonalsolution}).

It is  instructive to write down the KK monopole solution explicitly. The twisting in theory space is best performed in the string gauge, where the scalar vev points in the $T^3$ direction at infinity and our formulae for the $U_j$ of (\ref{Utheory}) are the appropriate ones. We begin by giving the explicit form of the diagonal solution in the $(v^l, \eta^l)$ vacuum in the string gauge. We define the polar coordinate components of the gauge field  by $A_k dx^k = A_r dr + A_\theta d \theta + A_\phi d\phi$, where $A_k$ are the Cartesian components. The string-gauge solution then reads:
\beqa
\label{stringdiagonal}
Q_j^{l, diag.} ~&=&~ v^l + i ~ e^{ i \: {\rm arg} \eta^l} ~{ f^l(r) \over r}~ T^3 ~, \nonumber \\
A_{j, r}^{l, diag.} ~&=& ~ 0~, \\
A_{j, \theta}^{l, diag.} ~&=&~ [ \chi^l(r) - 1 ]~e^{- i \phi T^3}~  T_2 ~e^{i \phi T^3}~,\nonumber \\
A_{j, \phi}^{l, diag.} ~&=&~ 4 \sin^2 {\theta \over 2} ~T^3  - [\chi^l(r)-1] \sin\theta~e^{- i \phi T^3}~  T_1 ~e^{i \phi T^3}~.\nonumber 
\eeqa
The functions $f^l, \chi^l$ appearing in (\ref{stringdiagonal}) are the same as  in eqn.~(\ref{solution}), but depend  on $\eta^l$ of eqn.~(\ref{vU}),  instead of $\eta$.
The large distance behavior of the solution in this gauge is transparent: since $[\chi^l -1]$ vanishes exponentially at large $r$, see (\ref{solution}), the only nonzero component of the gauge field is the first term in $A_\phi$ (giving rise to $B = 2 \sin\theta\: d\theta\: d\phi \:T^3$ and to the unit magnetic charge).

 To construct the KK monopoles, we now perform our $j$-dependent twists on the diagonal solution in the $v^l, \eta^l$ vacuum, using $(U_j^l)^{-1}$ of eqn.~(\ref{vU}). The result for the link fields is:
\beqa
\label{KKlink}
Q_{j}^{l,KK} = e^{i \; {\pi l \over N}} ~ v^l ~ U_1^{-1} \;  e^{i {2 \pi l \over N} T^3} \; U_1 + i ~e^{i \: \left[{\rm arg} (\eta^l) +{\pi l \over N} \right]  }~ {f^l(r) \over r} ~U_1^{-1} \;  e^{i {2 \pi l \over N} T^3} \; ~T^3~U_1 ~, 
\eeqa
%where the ``twisted" generators $T^A_{j,j^\prime; l}$ [$A = (a, 4)$   and $T^4 \equiv {\bf 1}$, as in %(\ref{linkdefs})] are:
%\beqa 
%\label{Tjj}
%T^A_{j,j^\prime; l} &=& U_1^{-1} ~e^{- i {2 \pi l (j -1)\over N}  ~T^a \hat{x}^a} ~T^A ~e^{ i {2 \pi  l %(j^\prime -1) \over N}  ~T^a \hat{x}^a}  ~   U_1  ~. 
%\eeqa
while  the gauge field of the $l$-th KK monopole is: 
\beqa
\label{KKgauge}
A_{j, r}^{l, KK} ~&=& ~ 0~, \nonumber \\
A_{j, \theta}^{l, KK} ~&=&~ [\chi^l(r) - 1]~U_1^{-1}~e^{- i \left[ \phi + {2 \pi l \over N} (j -1)\right] T^3}~  T_2 ~e^{i \left[ \phi + {2 \pi l \over N} (j -1)\right] T^3}~U_1~  \\
A_{j, \phi}^{l, KK} ~&=& ~4 \sin^2 {\theta \over 2}~ U_1^{-1} T^3 U_1~  - [\chi^l(r) - 1]\sin\theta~U_1^{-1}~ e^{- i \left[ \phi + {2 \pi l \over N} (j -1)\right] T^3}~  T_1 ~e^{i \left[\phi + {2 \pi l \over N} (j -1)\right] T^3}~U_1~.\nonumber 
\eeqa
 The above form explicitly shows the $j$-dependence of the KK monopole solution; in the string gauge, the link fields are still site-independent and the $j$-dependence enters solely through the theory space dependent shift of the azimuthal angle dependence of the gauge field of the monopole.
We stress again that (\ref{KKlink}, \ref{KKgauge}) is an exact solution of the classical equations of motion of the deconstructed theory at finite $N$,  in the $(v, \eta)$ vacuum; this is manifestly  true, as the KK monopole was constructed from  the diagonal solution (\ref{stringdiagonal}) using only the symmetries of the equations (\ref{eqngauge}-\ref{eqnlink}). It is easily seen, using (\ref{vU}), that at large distances $Q_{j; l}^{KK}$ asymptotes the $v,\eta$ vacuum (\ref{vacuum}). 

The form of the gauge potential (\ref{KKgauge}) of the KK monopole shows  that at large distances only the  first term in the azimuthal component of the gauge field survives, i.e. the magnetic field is now $B = 2 \sin\theta\: d\theta\: d\phi \: U_1^{-1}  T^3 U_1$. Recall that we have two possibilities for $U_1$---either it equals the identity or $i \sigma^2$.  Also note that the long range asymptotics of the link field is {\it not} twisted by $i \sigma_2$ (the effect of $i\sigma_2$ is taken into account in the different sign of $\eta$ in the relation between $v^l, \eta^l$ and $v, \eta$ in (\ref{vU}), as already mentioned). Thus,  solutions  twisted by $i \sigma_2$ have  magnetic charge opposite to that of the diagonal monopole they were obtained from, while solutions obtained via the trivial twist have the same charge. We will study further properties of the deconstructed KK monopoles in the following sections.
\footnote{A subset of the solutions (\ref{KKlink}, \ref{KKgauge})---the ones in the 
$v$, $\eta = 0$ vacuum,  with $v^l \sim v \cos (\pi l/N)$, $\eta^l \sim  v \sin (\pi \eta/N)$---have 
been found before \cite{KLEE}.  We thank Kimyeong Lee for pointing this out to us. }

\subsection{Quantum numbers and zero modes}

In the previous section, we obtained the KK monopole solutions,  constructed using $U^l_j$ of (\ref{Utheory}),  and the corresponding $v^l, \eta^l$, eqns.~(\ref{vU}). 
Since the KK monopoles are obtained by (global) gauge transformations of the known diagonal solution, 
all gauge invariant quantities characterizing the KK monopoles are simply equal to those of the diagonal solution in the $v^l, \eta^l$ vacuum. 

The Euclidean action of these solutions is given by   eqn.~(\ref{diagonalaction}) with $\eta$ replaced by $\eta^l$ (recall that $0 < l \le N$ for physically distinguishable solutions):
\beq
\label{action}
S_{KK}^{(l)} = {N \over g_3^2}~ 16 \pi  \left\vert \eta \cos {\pi l \over N} -  v \sin {\pi l \over N} \right\vert~,
\eeq 
There are two regions for $l$, near $l =0$ and   $l =  N$, where the action of the KK monopoles  remains finite at  zero lattice spacing and fixed $N/v$. (The action is then  proportional to the Coulomb branch vev $\eta$.)  There are two   solutions with the same action (\ref{action}) for any allowed nonzero value of $l$ (the KK monopoles   with and without an $i \sigma^2$ twist). The two towers differ by their other quantum numbers, however.

The magnetic charge of the KK monopole is  given by (\ref{magneticcharge}) evaluated on the solution. Since (\ref{KKgauge}, \ref{KKlink}) is obtained form (\ref{BPSdiagonalsolution}) in the $\eta^l, v^l$
vacuum by a symmetry transformation,  we may as well evaluate (\ref{magneticcharge}) on the diagonal solution in the $\eta^l, v^l$ vacuum. The only quantity that may change in  (\ref{magneticcharge}) is the chosen $\alpha$-th eigenvalue.
Since the spectra $\Omega_j (v^l, \eta^l)$ and $\Omega_j (v, \eta)$ are the same (their eigenvalues are gauge  invariant),  
the only two possibilities are that the chosen eigenvalues   remain  unchanged or are interchanged. 
Hence, the magnetic charge of the solution twisted by $U_{k}^l$ of eqn.~(\ref{Utheory}) is either equal or opposite that of the diagonal solution one starts with. 
The two eigenvalues of $\Omega_j$ are interchanged only for solutions obtained by applying the $i \sigma^2$ transformation (\ref{U1}).  
Equivalently, the change of magnetic charge for the solutions obtained by the $i \sigma^2$ twist can be seen by simply noting that the scalar field asymptotics are not changed by this twist (this is compensated by the already noted change of sign of $\eta$ in the relation between $v^l, \eta^l$ and $\eta,v$, (\ref{vU}), for $U_1$ of the second type in (\ref{U1})) but the sign of the magnetic field changes. 

To summarize the KK monopole spectrum:  for every self-dual diagonal monopole of magnetic charge $+1$ there are  $N-1$ self-dual KK monopoles with magnetic charges $+1$ and $N-1$ self-dual
KK monopoles with magnetic charge $-1$ (the ones twisted by $i \sigma^2$). 
A similar tower of KK-anti-monopoles of charges $\pm 1$ will of course result starting from the anti-self-dual diagonal antimonopole of charge $-1$.

Solutions of the continuum 4d theory compactified on  a circle are also characterized by the topological charge
(not necessarily integer if a Wilson line is turned on; for a topological classification of finite action solutions on $R^3 \times S^1$, see \cite{Gross:1980br}).
To define the topological charge  we need  an expression for the electric field. The field $\vec{{\cal E}}_j(x)$ at the $j$-th site  should be a hermitean group algebra element, transforming homogeneously under gauge transformations at the $j$-th site. A natural definition is:
\beq
\label{electric} 
\vec{{\cal E}}_j(x) =   i ~\vec{D} ~ {\cal U}_j  \cdot {\cal U}_j^{-1} ~
\equiv~  i~ \vec{\nabla} ~{\cal U}_j \cdot {\cal{U}}_j^{-1}  - {1\over 2} \left( \vec{A}_j - {\cal U}_j \vec{A}_{j+1} {\cal U}_j^{-1} \right)~,
\eeq
where ${\cal U}_j$  are the unitary link fields, defined in section 3.1. 
Under gauge transformations $\vec{{\cal E}}_j(x) \rightarrow U_j \vec{{\cal E}}_j(x) U_j^{-1}$, as required. Then, the gauge invariant  and local, both in real and ``theory" space, expression: 
\beq
\label{Q}
Q  = {1 \over 8 \pi^2}~ \sum\limits_{j = 1}^N \int d^3 x~ {\rm tr} ~\vec{{\cal E}}_j (x) \cdot \vec{B}_j(x)
\eeq
evidently has the correct continuum limit. As usual, it suffices to find  the topological charge for the diagonal monopoles; the KK monopoles   have the same charge  expressed in terms of $v^l, \eta^l$.
The shortest route to computing $Q$ is to find the expressions for $\vec{{\cal E}}$ and $\vec{B}$ for the diagonal solution. After a straightforward computation (for the sake of conciseness, we give the explicit formulae in the footnote\footnote{More precisely, in terms of $f(r \eta)$ and $\chi(r \eta)$ defined after eqn.~(\ref{BPSdiagonalsolution}), and $F \equiv f/(r v)$, we find
${\cal E}_j^m = (1+ F^2/4)^{-1} \left[ \epsilon^{m b c} n^b T^c F^2 ~{ 1 - \chi \over 4 r}  +  T^m F ~{1 - \chi \over 2 r} + n^m n^a T^a  \left(F^\prime - F~{1-\chi \over 2 r} \right)\right]$, while $B_j^m = T^m \left[ \left({1 + \chi \over  r}\right)^\prime + {1 + \chi \over r^2 }\right] - n^m T^a n^a \left[ {1\over 2}\left({1+\chi \over r}\right)^2 + \left({1 + \chi\over r}\right)^\prime - {1 +\chi\over r^2} \right]
$. Since the $F^2$ term in the electric field scales like the square of the lattice spacing, it is evident, upon comparing with the explicit solution (\ref{solution}), that  self-duality (\ref{Q1}) holds near the continuum limit.}) it is easy to deduce that:
\beq
\label{Q1}
\vec{{\cal E}}_j (x) = {1\over v}~ \vec{B_j} (x) + {\cal O}\left({1\over v^2}\right)~,
\eeq
which shows that the  self-duality of the solution holds in the continuum limit. Then, 
it follows from (\ref{Q1}) that, in the same limit, the topological charge (\ref{Q}) of the solution  is given by
(to evaluate the integral, compare with eqn.~(\ref{diagonalaction})):  
\beq
\label{Q2}
Q \simeq {N \over 8 \pi^2 v} ~\int d^3 x~ {\rm tr} ~\vec{B}_{j = 1}^2 ~\simeq ~{N \over   \pi v}~ ~|\eta|~ =  ~R ~|\eta|~.
\eeq
The topological charge of the KK monopoles is given by the same expression, but with $v, \eta$ replaced by $v^l, \eta^l$ of (\ref{vU}) [in the continuum limit, where the r.h.s. of (\ref{Q2}) is valid].

Turning to the zero modes, we note that they, too, can be obtained from the zero modes of the diagonal solution in the $v^l, \eta^l$ vacuum after applying the appropriate $U(2)^N$ transformation. The diagonal solution has four bosonic zero modes, three associated with its position in the 3d Euclidean space and one with the orientation in the unbroken $U(1)$ gauge group. The diagonal solution also has four fermionic zero modes, as is known, for example, from the studies \cite{Seiberg:1996nz} of  Seiberg-Witten dynamics compactified to three
dimensions (hence the monopoles do not contribute to the superpotential).  

Finally, we note that the KK monopoles do not preserve any of the explicit supersymmetry of the 
$SU(2)^N$ theory (this is easy to see by considering the explicit $N=2$ 3d  supersymmetry variations of the fields). Despite completely breaking the finite-$N$ supersymmetry, the KK monopoles are topologically stable solutions. At infinite $N$, the KK monopoles become BPS saturated and preserve half of the continuum limit 4d $N=2$ supersymmetry.

\section{Deconstructing the 4d instanton}

We first recall  the significance of the continuum KK monopole solution as a constituent of the 4d instanton. 
In the continuum theory \cite{Lee:1997vp},\cite{Lee:1998bb},\cite{Kraan:1998pm},\cite{Kraan:1998kp}, it was argued that a solution obtained by superposing the  diagonal monopole  with an appropriate KK monopole solution  is, in fact, the  4d instanton solution; in other words, the diagonal and KK monopole solutions can be viewed as  ``constituents" of the 4d instanton.

We can see how this works out in deconstruction by considering, for example, the formulae for the action of the  diagonal and an appropriate KK  monopole. Consider the KK monopole with $l = -1$ (a case of special interest, as we will see below) twisted by $i \sigma^2$, i.e. with opposite sign  of $\eta$ in the relation between $\eta, v$ and $\eta^l, v^l$ (\ref{vU}). The action of this solution is then given, taking {\it real}  $\eta < 1/R$, by (\ref{action}) expanded for $l \ll N$: 
\beq
\label{SKK}
S_{KK, l = -1, \sigma^2} = {N \over g_3^2} ~16 \pi | \eta^l| \simeq {N \over g^2_3} ~16 \pi \left({1 \over R} - \eta \right) = {2 \pi R \over g^2_4} ~16 \pi \left({1 \over R} - \eta \right) ~. 
\eeq
Recall also that the magnetic charge of this solution is $-1$. 
Now, consider also the diagonal solution (\ref{solution}) with action (\ref{diagonalaction}) and magnetic charge  $+1$. 

If we superpose these two solutions, placing them sufficiently far apart, we obtain a field configuration, which we expect to be an approximate minimum of the action, even for finite $N$. The action of this field configuration will be approximately equal to the sum of the two actions, eqn.~(\ref{diagonalaction}) and (\ref{SKK}) above:
\beq
\label{Ssum}
S \simeq S_{diag} + S_{KK, l =-1, \sigma^2 } ~ \simeq~  {N \over g_3^2} ~ 16 \pi \left[  \eta + \left({1 \over R} - \eta\right) \right] ~=~  {N \over g_3^2} ~ {16 \pi     \over R} ~ =~ {32 \pi^2 \over g_4^2}~  \equiv ~ { 8 \pi^2 \over {\tilde{g}}_4^2} ~.
\eeq
The action thus approaches the action of the 4d instanton (in the last equality above, we recall the footnote from section 3.1: to achieve conventional normalization, we rescale $g_4 = 2 {\tilde g}_4$).  

It is important to note that the action (\ref{Ssum}), when expressed in terms of $g_4$ is independent of the radius of compactification.
Moreover, its independence also on $\eta$   suggests that the monopole--KK monopole configuration will  persist as   (approximate) solution also at the origin of the Coulomb branch. 
The simultaneous $\eta$ and $R$ independence  of the action (\ref{Ssum}) is only possible on the branch where the imaginary part of $\eta$ vanishes.  It is easy to see, using our formulae for the actions, that if any vev on the 4d Coulomb branch was turned on (e.g. $\eta$ is not real), the combined action would diverge at infinite $R$, for nonvanishing vev, as  the action would then be  $32 \pi^2 g_4^{-2} (\sqrt{1 + R^2 |\eta|^2} + R |\eta|)$; the same comment applies to the topological charge below.

Similarly, we can also argue that (in the string  gauge) the topological charge of the diagonal monopole and KK monopole configuration is approximately equal to the sum of the topological charges. Using eqn.~(\ref{Q2}) for the diagonal  monopole and the same expression, but with $\eta \rightarrow 1/R - \eta$, for KK monopole, we find that at large $N$: 
\beq
\label{Qsum}
Q \simeq Q_{diag} + Q_{KK, l =-1, \sigma^2 } \simeq R \eta + R \left({1\over R } - \eta\right) = 1~.
\eeq
In the large-$N$ limit, both solutions become BPS saturated, preserving half of the continuum limit enhanced 4d $N=2$ supersymmetry. Thus, in the large-$N$ limit,  we can superimpose the solutions   for arbitrary separations and can appeal to the continuum analysis of  ref.~\cite{Lee:1998bb}, where it was explicitly shown that the field of the 4d instanton is reproduced in the appropriate limit.  We should also note that the number of bosonic and fermionic zero modes---four per solution---also matches the eight bosonic and fermionic zero modes of the instanton in the 4d $N=2$ theory. The eight (approximate  at finite $N$) fermionic zero modes of the superimposed monopole--KK monopole configuration   become exact zero modes in the continuum limit, giving rise to the 4d $U(1)_R$-violating 't Hooft vertex (we do not give a detailed analysis, which is bound to be rather complicated, but  appeal, as before, to the continuum study). 

The fact that the diagonal and KK monopole solutions are only ``approximately BPS" (at large-$N$)   complicates  the explicit study of the superimposed solutions at finite $N$; for example, we expect that there is a $\sim 1/N$ force between them. 

If we dimensionally uplift the theory to deconstruct $N=1$ 5d super Yang-Mills theory, the KK monopole solutions we study become magnetically charged static stringlike configurations. In the classical continuum limit one can similarly argue that the monopole and KK monopole strings make up a 5d gauge soliton with unit topological charge (\ref{Q}), which is now the zeroth component of the 5d topological current. The energy  of the configuration of the two superimposed strings is then given by the 5d analogue of (\ref{Ssum}) and remains finite even in the decompactification limit.  An analysis of the quantization of the noncompact size modulus of the gauge soliton (classically, solitons of arbitrary size are allowed) and of its possible stabilization, which clearly depends on the UV completion, i.e. finite $N$,  will not be attempted here.
 
\section{Conclusions and outlook}

In this paper, we studied nonperturbative effects in the deconstruction of 4d  and 5d supersymmetric theories with 8 supercharges. In the 3d $\rightarrow$ 4d case, we completed the study (which began by an investigation of deconstructed theory anomalies in perturbation theory in \cite{Giedt:2003xr}) of deconstructed theory anomalies by explicitly finding the nonperturbative objects on the lattice that lead to anomalous charge violation in the continuum.    

While it is comforting, at least to the author, to know  precisely how deconstructing  a 4d ${\cal{N}}=2$ theory reproduces the known nonperturbative properties of the continuum theory, this does not seem to add anything substantially new to our understanding of Seiberg-Witten theory. Thus our results should be viewed as contributing to a better understanding of how deconstruction works,  at least in cases where the continuum theory dynamics is known.

The potentially more interesting  applications of deconstruction are  in the study of higher-dimensional theories, with or without supersymmetry. Continuum 5d theory gauge solitons (uplifted 4d  instantons) have been conjectured to play an important role in the nonperturbative dynamics of the theory. Studying their dynamics in the continuum theory is complicated, however, since the soliton  properties   are   sensitive to the UV completion of the theory (e.g., to various higher-dimensional operators).
 In particular, the  interpretation of the soliton states in the continuum quantum theory is unclear, since it is  difficult to quantize their noncompact zero modes in the absence of a UV completion (string theory can provide one possible completion).
We have given  a construction of nonperturbative objects in the 5d theory---the constituents of  5d gauge solitons---in the framework of deconstruction, a gauge invariant UV completion of the theory. It would be interesting to learn more about the properties of these states in   deconstruction. 

One would also like to study the appearance of $S$-duality and other symmetries of the continuum limit theory at finite $N$ (one  observation in this regard is that the {\it finite $N$} spectrum of the magnetically charged strings in the ${\cal{N}} =1$ quiver theory, the Minkowski analog of eqn.~(\ref{action}),  is identical, up to   $g^2 \rightarrow {1\over g^2}$, to the spectrum of $W^{\pm}$ bosons, given in  ref.~\cite{Csaki:2001zx}).

Another possible direction where the  investigation of nonperturbative objects (black holes) via deconstruction might be of interest is  in deconstructed theories of gravity 
\cite{Bander:2001qk},   \cite{Arkani-Hamed:2002sp}, \cite{Jejjala:2002we}, \cite{Arkani-Hamed:2003vb}. Whether the methods for constructing nonperturbative solutions described here will turn out to be useful remains to be seen; studying this might be easier in deconstructed versions of supergravity.

\section{Acknowldegments}

The author would like to thank   Moshe Rozali for collaboration at the initial stage of this work and   Joel Giedt for comments on the manuscript. This work is supported by the National Science and Engineering Research Council of Canada.

 \end{document}